\documentclass{article}
\usepackage{booktabs}
\usepackage[table,xcdraw]{xcolor}
\usepackage{spconf,amsmath,graphicx}
\usepackage{multirow}
\usepackage{hyperref}
\usepackage{marvosym}

\definecolor{darkgreen}{rgb}{0, 0.5, 0}
\definecolor{darkred}{rgb}{0.5, 0, 0}
\definecolor{darkpurple}{rgb}{0.4, 0, 0.4}
\definecolor{darkgray}{rgb}{0.50, 0.50, 0.50} 

\newcommand{\bt}[1]{\textcolor{blue}{#1}}
\newcommand{\gt}[1]{\textcolor{darkgreen}{#1}}
\newcommand{\rt}[1]{\textcolor{darkred}{#1}}

\newcommand{\grt}[1]{\textcolor{darkgray}{#1}}
\newcommand{\squeezeup}{\vspace{-2.0mm}}

\title{Joint Audio and Speech Understanding \vspace{-2mm}}

\name{
Yuan Gong\textsuperscript{\href{mailto:yuangong@mit.edu}{\Letter}}$^1$, Alexander H. Liu$^1$, Hongyin Luo$^1$, Leonid Karlinsky$^2$, James Glass$^1$  \vspace{-4mm}}
\address{
$^1$MIT CSAIL, USA \quad\quad $^2$MIT-IBM Watson AI Lab, USA \vspace{-3mm}
\thanks{
This research is supported by the MIT-IBM Watson AI Lab. Code, dataset, and pretrained models are at \href{https://github.com/yuangongnd/ltu}{\color{blue}{github.com/yuangongnd/ltu}}.}
}

\begin{document}

\maketitle
\begin{abstract}

Humans are surrounded by audio signals that include both speech and non-speech sounds. The recognition and understanding of speech and non-speech audio events, along with a profound comprehension of the relationship between them, constitute fundamental cognitive capabilities. For the first time, we build a machine learning model, called \texttt{LTU-AS}, that has a conceptually similar universal audio perception and advanced reasoning ability. Specifically, by integrating Whisper~\cite{radford2022robust} as a perception module and LLaMA~\cite{touvron2023llama} as a reasoning module, \texttt{LTU-AS} can \emph{simultaneously} recognize and \emph{jointly} understand spoken text, speech paralinguistics, and non-speech audio events - almost everything perceivable from audio signals.
\end{abstract}

\squeezeup\squeezeup\squeezeup
\section{Introduction}
\squeezeup
\label{sec:intro}

Humans live in a multifarious environment of audio signals, encompassing both speech and a wide variety of non-speech sounds. The ability to accurately discern, interpret, and integrate these speech and non-speech audio elements, in conjunction with a profound understanding of the interrelationships they entail, represents a fundamental cognitive capability of humans. 
When we hear ``watch out!'' and a car horn simultaneously, we can infer the danger. If we hear birds chirping and someone says ``that's a rare one," we know there is an unusual bird nearby. Understanding music usually requires paying attention to both the lyrics and the melody.

\begin{figure}[!t]
\centering
\includegraphics[width=6.85cm]{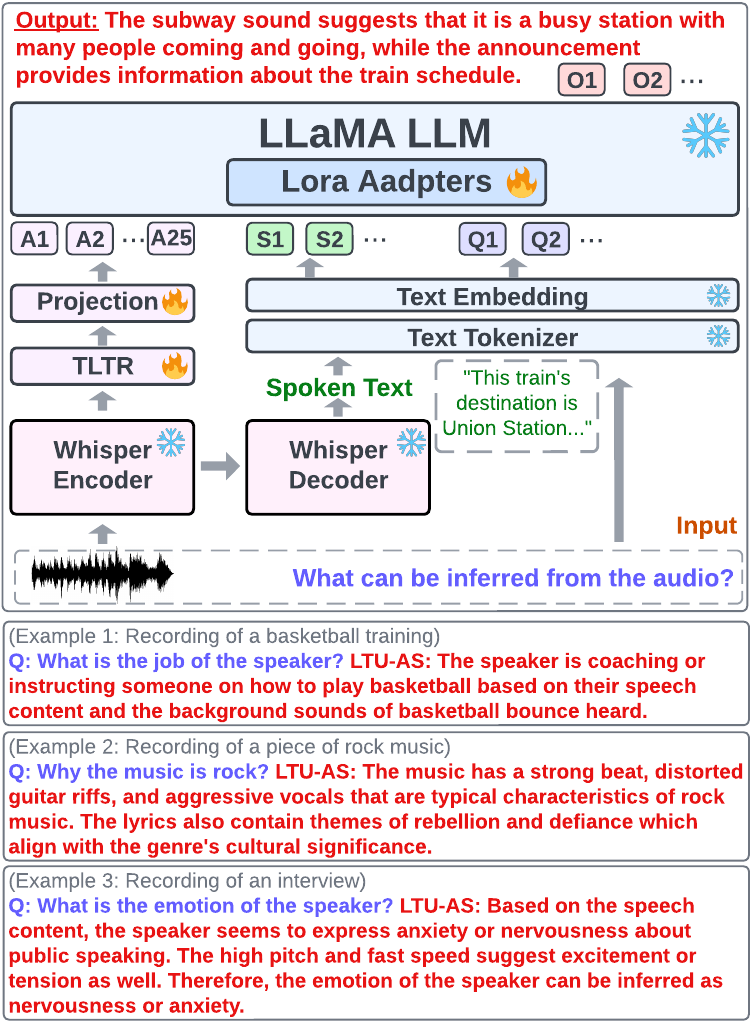}
\squeezeup\vspace{-1mm}
\caption{Illustration of the \texttt{LTU-AS} model and real samples showing its \emph{joint} audio and speech understanding ability.}
\label{fig:ltue}
\squeezeup\squeezeup
\end{figure}

However, most existing machine learning models can only recognize either speech or audio events. Further, while being strong in audio or speech perception, these models possess limited reasoning and understanding capabilities. This motivates us to build a \emph{joint audio and speech understanding} model that is able to simultaneously recognize and jointly understand speech and audio events. Particularly, as shown in Figure~\ref{fig:ltue}, our model integrates pretrained Whisper~\cite{radford2022robust} automatic speech recognizer (ASR) and a time and layer-wise Transformer (TLTR)~\cite{gong2023whisper} as the perception module and LLaMA~\cite{touvron2023llama} large language model (LLM) as the reasoning module. In addition, we formulate the training data as  (audio, question, answer) (AQA) tuples, which allows us to combine 13 audio and speech datasets of various tasks with different label sets into a single 9.6M \texttt{Open-ASQA} dataset, among which 6.9 million data are open-ended AQA tuples generated by GPT~\cite{brown2020language} with \emph{audio instruction generation}~\cite{gong2023listen}. We call our model \texttt{LTU-AS} (listen to, think of, and understand audio and speech). Performance-wise, we show \texttt{LTU-AS} is strong on all audio/speech tasks. But more importantly, as shown in Fig.~\ref{fig:ltue} and Table~\ref{tab:open_res}, \texttt{LTU-AS} can answer free-form open-ended questions about the audio and speech with an instruction following rate over 95\% (evaluated by GPT-4), and exhibits emerging joint audio and speech reasoning ability.

\vspace{2.mm}
\noindent{\textbf{Related Work:}} \texttt{LTU-AS} substantially improves the recent audio large language model LTU~\cite{gong2023listen} that only understands non-speech audio. Particularly, \texttt{LTU-AS} adopts Whisper~\cite{radford2022robust} and TLTR~\cite{gong2023whisper} as the audio encoder instead of the AST~\cite{gong21b_interspeech} audio encoder in LTU. This change enables \texttt{LTU-AS} to recognize both speech and audio events. We also augment the LTU OpenAQA-5M dataset with 4 million speech and audio/speech understanding AQAs in creating the 9.6M \texttt{Open-ASQA} dataset. There are a few recent efforts on joint audio and speech recogntion~\cite{dinkel2022unikw,moritz2020all,narisetty2022joint,turian2022hear} but none of them exhibit advanced joint reasoning ability. Other recent audio LLMs~\cite{gao2022wavprompt,huang2023audiogpt,chang2023speechprompt,zhang2023speechgpt} primarily focus on only speech. To the best of our knowledge, \texttt{LTU-AS} is the first joint audio and speech understanding model.

\section{LTU-AS Model Architecture}
\label{sec:arc}
\subsection{Design Overview}
\squeezeup

The architecture of \texttt{LTU-AS} is depicted in Fig.~\ref{fig:ltue}. The system input is a pair of audio and question in natural language form. The audio is first input to the Whisper audio encoder. Then, the output of the Whisper encoder is fed to the Whisper decoder to transcribe it to \emph{discrete} spoken text (if there is no speech, then the output of the decoder will be empty, which is as expected). Meanwhile, we feed the output of all 32 Whisper encoder intermediate layers to an AudioSet-pretrained Time and Layer-Wise Transformer (TLTR)~\cite{gong2023whisper} to encode ``soft'' audio events and speech paralinguistic information, and then project to a series of \emph{continuous} audio tokens $\{A\}$ with a linear layer. 

During training, the entire Whisper model is frozen. Only the TLTR model and projection layer are trainable. This design is due to a few reasons: First, training a large language model as an automatic speech recognizer (ASR) can be very expensive but the benefit is unclear~\cite{zhang2023speechgpt}, we thus choose to freeze the entire Whisper model to inherit its strong and robust ASR ability. Second, although the Whisper encoder encodes rich audio events and speech paralinguistic information~\cite{gong2023whisper,feng2023peft,xu2023understanding}, it encodes information in the representations of different layers. Since we anticipate \texttt{LTU-AS} being a universal perception model, we use the TLTR model to apply attention mechanisms over both time and layers. 

The key advantage of this setting is that the audio is encoded to both text and continuous tokens, so both linguistic and non-linguistic information are kept. We then tokenize and embed the spoken text and input question to a sequence of text tokens $\{S\}$ and $\{Q\}$, respectively. Finally, we concatenate and input $\{A\}$, $\{S\}$, and $\{Q\}$ to the LLaMA LLM. Due to the computational limit, we trim the length of the audio token $\{A\}$ to 25 (corresponding to 10 seconds of audio), but allow $\{S\}$ and $\{Q\}$ to be of variable length. 

\squeezeup\squeezeup
\subsection{Audio Encoder}
\textbf{Whisper}~\cite{radford2022robust} is a recently proposed robust ASR model that features a standard Transformer~\cite{vaswani2017attention}-based encoder-decoder architecture trained with a massive 680k hour labeled speech corpus recorded in diverse conditions. Notably, it was found that the Whisper encoder features not only encode linguistic information, but also encode rich general background sound information~\cite{gong2023whisper} and paralinguistic and other information (e.g., emotion~\cite{feng2023peft} and language development~\cite{xu2023understanding}). In this paper, we use the Whisper-large model whose encoder and decoder are both 32-layer, 1280-dimensional Transformer networks.

\noindent{\textbf{Time and Layer-Wise Transformer (TLTR)}}: We use the AudioSet pretrained TLTR for Whisper proposed in~\cite{gong2023whisper}, originally for audio event detection. We empirically find there is no need to pretrain it further on speech classification tasks before training together with \texttt{LTU-AS}. Whisper and TLTR pool the audio with a factor of 40, i.e., for each 10-second audio (1000 frames), the length of the TLTR output is 25 (2.5Hz).

\noindent{\textbf{Projection Layer}}: We use a single linear layer to project the TLTR output from 1280-dimensional to 4096-dimensional to match the embedding dimension of the LLaMA LLM. 

\squeezeup\squeezeup
\subsection{LLaMA Large Language Model}

We use the LLaMA-7B LLM~\cite{touvron2023llama} with Vicuna~\cite{vicuna2023} instruction following for fine-tuning. To mitigate catastrophic forgetting~\cite{goodfellow2013empirical} and save computation, we freeze the entire LLaMA model and adopt Low-rank Adaptation \cite{hu2021lora} (LoRA), which introduces a small set of auxiliary learnable weights on top of the pre-trained LLaMA model. Specifically, we inject LoRA adapters (rank=8 and $\alpha$=16) to the projection layers for all keys and queries in all LLaMA self-attention layers~\cite{vaswani2017attention}.


\noindent{\textbf{Training Objective:}} As an audio LLM, \texttt{LTU-AS} is trained on the next token prediction task conditioning on the past tokens and the reference audio, i.e., maximizing $P(O_t \mid O_{1:t-1}, A, S, Q)$, through cross-entropy for all $1<t \leq T$ given the tokenized ground truth text sequence (i.e., output) $O_{1:T}$ and the reference audio token $A$, spoken text $S$, and question $Q$. This training objective allows us to unify nearly all audio and speech tasks except audio/speech generation into a single training framework.

\noindent{\textbf{Generation Setting:}} We use a plain generation setting of Temperature$=$0.1, Top K$=$500, and Top P$=$0.95 with a repetition penalty of 1.1~\cite{fan2018hierarchical,keskar2019ctrl} for all tasks. 

\noindent{\textbf{Model Parameters:}} As a LLM, \texttt{LTU-AS} has about 8.5 billion parameters. However, only 49 million parameters are actually trainable (40M for TLTR, 4.2M for LoRA adapters, and 5M for the projection layer), which is only about 0.6\% of the total number of parameters. This significantly lowers the computation requirement to train \texttt{LTU-AS}. Practically, \texttt{LTU-AS} is trained on 4$\times$ A6000 GPUs for about 80 hours.

\squeezeup
\section{The Open-ASQA Dataset}
\label{sec:data}
\squeezeup

We aim to build \texttt{LTU-AS} to address a wide range of open-ended audio and speech tasks, and understand the audio and speech jointly. To achieve this objective, we need a training dataset to provide such joint audio and speech supervision. Unfortunately, there is no existing dataset that meets our needs. The closest one is the OpenAQA dataset used to train LTU~\cite{gong2023listen}, which is an audio question-answering dataset consisting of 1.9 million closed-ended and 3.7 million open-ended AQAs. However, OpenAQA lacks speech related, and joint audio-speech questions. Therefore, on the basis of OpenAQA-5M, we add an additional 2.7 million speech-related AQAs (0.9 million closed-ended and 1.8 million open-ended) and 1.2 million joint audio and speech AQAs (almost all open-ended), and build a new 9.6M \texttt{Open-ASQA} dataset. Note that we do not collect new audio and speech data, but instead relabel 13 existing public datasets summarized in Table~\ref{tab:openasqa}. For all these datasets, we only include data marked as training and validation samples and exclude any data marked as test or evaluation.

As with OpenAQA, all \texttt{Open-ASQA} samples are formatted as (audio, question, answer) tuples, where ``audio'' and ``question'' are the model inputs, and ``answer'' is the ground truth label. By unifying all training samples in this format, we not only map all labels to a semantic space, but are also able to train \texttt{LTU-AS} with a variety of different tasks easily.

\squeezeup
\subsection{Closed-Ended AQA Generation}

For each task and dataset, we paraphrase the question (e.g., ``What is the audio event'') with GPT-3.5-Turbo assistance to generate a diverse question set, so \texttt{LTU-AS} won't overfit to a specific question for a task. However, the answers are generated with a rule-based algorithm based on the original label of the dataset, and thus have a fixed format. We thus call such AQAs closed-ended AQAs. The upper section of Table~\ref{tab:sample} shows samples of closed-ended AQA pairs. 

\textbf{Closed-Ended Audio AQA}: Closed-ended audio questions are from the original OpenAQA dataset, which consists of 1.9 million AQAs about the audio event labels, acoustic features, audio captioning, and audio temporal analysis. The audio tracks are from 8 audio datasets. Please refer to Table~\ref{tab:openasqa} and~\cite{gong2023listen} for more details. 

\textbf{Closed-Ended Speech AQA}: We created 941k closed-ended speech AQAs based on 4 commonly used speech datasets. The first category of questions asks the original labels of the datasets, e.g., speaker emotion/gender for IEMOCAP~\cite{busso2008iemocap}, speaker emotion and sentiment score for MOSEI~\cite{zadeh2018multimodal}, speaker gender for LibriTTS~\cite{zen2019libritts}, and speaker age and gender for VoxCeleb2~\cite{chung2018voxceleb2,hechmi2021voxceleb}. In addition to these original labels, we further annotate the speaker style of speech speed (computed with Whisper time stamps), pitch, and energy (computed with librosa~\cite{mcfee2015librosa}), and generate AQAs asking the speaker style. Finally, we also mix in about 150k ASR AQAs that have questions asking about the spoken text, and the answers are the transcriptions. Note that since \texttt{LTU-AS} has an internal Whisper model feeding the transcribed text to LLaMA, the ASR task is no more than an identity mapping for \texttt{LTU-AS}, which is fundamentally different from SpeechGPT~\cite{zhang2023speechgpt}. We include ASR AQAs just to guide the model following ASR instructions.

\textbf{Closed-Ended Joint Audio and Speech AQA}: Most joint audio and speech AQAs in this project are open-ended. The only 93k closed-ended joint audio and speech AQAs are of the music genre prediction task on the FMA~\cite{defferrard2017fma} dataset, which requires \texttt{LTU-AS} to consider both lyrics (text) and acoustic information to make the prediction.

\begin{table}[t]
\footnotesize
\centering
\squeezeup
\caption{Sample closed- and open-ended training QA pairs. Open-ended QA pairs are generated by \emph{Audio Instruction Generation} (AIG) with the shown GPT prompt and input. \\ *GPT prompt is shortened due to space limitation.}
\label{tab:sample}
\begin{tabular}{p{8cm}}
\toprule
\textbf{Closed-Ended QA Samples}                                                                                                                                                                 \\ \midrule
\bt{Q: Identify the sound from the clip?} \rt{A: Dog, Animal.}\\
\bt{Q: Write an audio caption describing the sound?} \\
\rt{A: An engine revving and then tires squealing.} \\
\bt{Q: What emotion does the speaker convey?} \\ \rt{A: Happy, with a sentiment score of 2 (positive).}  \\
\bt{Q: Can you determine the music genre?} \rt{A: Folk, Pop.} \\ \midrule\midrule
\textbf{Open-Ended QA Samples}  \\ \midrule
\textbf{GPT Prompt*}: Based on the following audio/speech, generate 10 different types of complex open-ended questions that require step-by-step thinking, and corresponding answers. Questions can be e.g., How are speech content and background sounds related? ... etc. Format each QA pair in a single line as a JSON dictionary. \grt{(varies for each task/dataset)} \\ \midrule
\textbf{GPT-Input}: In the recording, background sound of domestic animals, dog and speech of ``Get him going, get him going with it.'' is heard.  \\
\textbf{GPT-Output}: \bt{Q: What can be inferred from the audio?} \rt{A: It can be inferred that the speech is addressing the dog in the background.}  \\ \midrule
\textbf{GPT-Input}: In the recording, background sound of explosion and speech of ``Flames engulfed the tank and began to heat the propane inside. Twenty volunteer firefighters...'' is heard.  \\
\textbf{GPT-Output}: \bt{Q: Why do the speech content and explosion sound create a tense atmosphere?} \rt{A: The speech content describes a tragic event and the explosion sound adds intensity and suspense to the narrative.} \\ \midrule
\textbf{GPT-Input}: Speech: ``Hi, my name is Jamie and I'm a Master Sales Consultant here at Northline Volkswagen...''; Speaker gender: Male; Pitch: low; Volume: very high; Speed: very slow; Emotion: happy. \\
\textbf{GPT-Output}: \bt{Q: What can we infer from the speech content and emotion?} \rt{A: The speaker seems happy to introduce himself and his company, and he is delivering his words in a slow and clear manner to make sure everyone understands what he is saying.} \\ \midrule
\textbf{GPT-Input}: Music genre: Folk; Music Lyrics: ``The man of his dreams is all that he seems Flowers in sunshine...'' \\
\textbf{GPT-Output}: \bt{Q: What mood does this music convey?} \\\rt{A: The music conveys a cheerful and relaxed mood because of the lively melody and lyrics about flowers and sunshine.}\\ \bottomrule
\end{tabular}
\squeezeup
\end{table}

\begin{table*}[t]
\footnotesize
\centering
\squeezeup\squeezeup\squeezeup
\caption{The statistics of the 9.6-million \texttt{Open-ASQA} dataset. ``x'' denotes the corresponding label is used.}
\label{tab:openasqa}
\begin{tabular}{@{}lccccccccccc@{}}
\toprule
Dataset     & \begin{tabular}[c]{@{}c@{}}Audio\\ Event\end{tabular} & \begin{tabular}[c]{@{}c@{}}Audio\\ Caption\end{tabular} & \begin{tabular}[c]{@{}c@{}}Spoken\\ Text*\end{tabular} & \begin{tabular}[c]{@{}c@{}}Speaker\\ Gender\end{tabular} & \begin{tabular}[c]{@{}c@{}}Speaker\\ Age\end{tabular} & \begin{tabular}[c]{@{}c@{}}Speech\\ Style\end{tabular} & \begin{tabular}[c]{@{}c@{}}Speaker\\ Emotion\end{tabular} & \begin{tabular}[c]{@{}c@{}}Music\\ Genre\end{tabular} & \begin{tabular}[c]{@{}c@{}}\# Audio\\ Clips\end{tabular} & \begin{tabular}[c]{@{}c@{}}\# Closed-\\ Ended QAs\end{tabular} & \begin{tabular}[c]{@{}c@{}}\# Open-\\ Ended QAs\end{tabular} \\ \midrule
\multicolumn{12}{l}{{\color[HTML]{9B9B9B} \textit{\textbf{Audio Datasets (OpenAQA)}}~\cite{gong2023listen}}}   \\ \midrule
AS-Strong~\cite{hershey2021benefit}   & x                                                     & x                                                       & x                                                      & x                                                        & -                                                     & -                                                      & -                                                         & -                                                     & 102k                                                     & 683k                                                          & 901k                                                        \\
AudioSet~\cite{gemmeke2017audio}    & x                                                     & -                                                       & x                                                      & x                                                        & -                                                     & -                                                      & -                                                         & x                                                     & 500k                                                     & 538k                                                          & 184k                                                        \\
VGGSound~\cite{chen2020vggsound}    & x                                                     & -                                                       & x                                                      & x                                                        & -                                                     & -                                                      & -                                                         & x                                                     & 184k                                                     & 367k                                                          & 907k                                                        \\
FSD50K~\cite{fonseca2021fsd50k}      & x                                                     & -                                                       & x                                                      & x                                                        & -                                                     & -                                                      & -                                                         & x                                                     & 41k                                                      & 82k                                                           & 403k                                                        \\
AudioCaps~\cite{kim2019audiocaps}   & x                                                     & x                                                       & x                                                      & x                                                        & -                                                     & -                                                      & -                                                         & x                                                     & 46k                                                      & 97k                                                           & 478k                                                        \\
FreeSound~\cite{freesound}   & -                                                     & x                                                       & x                                                      & -                                                        & -                                                     & -                                                      & -                                                         & -                                                     & 91k                                                      & 91k                                                           & 791k                                                        \\
Clotho~\cite{Drossos_2019_dcase}      & -                                                     & x                                                       & x                                                      & -                                                        & -                                                     & -                                                      & -                                                         & -                                                     & 5k                                                       & 48k                                                           & 89k                                                         \\
Sound Bible~\cite{soundbible} & -                                                     & x                                                       & x                                                      & -                                                        & -                                                     & -                                                      & -                                                         & -                                                     & 1.2k                                                     & 12k                                                           & 10k                                                         \\
\rowcolor[HTML]{EFEFEF} 
Sum         &                                                       &                                                         &                                                        &                                                          &                                                       &                                                        &                                                           &                                                       & 845k                                                     & 1,918k                                                        & 3,763k                                                      \\ \midrule
\multicolumn{12}{l}{{\color[HTML]{9B9B9B} \textit{\textbf{Speech Datasets}}}}                                                                                                                                                                                                                                                                                                                                                                                                                                                                                                                                                                                                   \\ \midrule
IEMOCAP~\cite{busso2008iemocap}     & -                                                     & -                                                       & x                                                      & x                                                        & -                                                     & x                                                      & x                                                         & -                                                     & 4.3k                                                     & 26k                                                           & 83k                                                         \\
LibriTTS~\cite{zen2019libritts}    & -                                                     & -                                                       & x                                                      & x                                                        & -                                                     & x                                                      & -                                                         & -                                                     & 22k                                                      & 167k                                                          & 418k                                                        \\
VoxCeleb2~\cite{chung2018voxceleb2}    & -                                                     & -                                                       & x                                                      & x                                                        & x                                                     & x                                                      & -                                                         & -                                                     & 107k                                                     & 194k                                                          & 926k                                                        \\
MOSEI~\cite{zadeh2018multimodal}       & -                                                     & -                                                       & x                                                      & -                                                        & -                                                     & x                                                      & x                                                         & -                                                     & 18k                                                      & 554k                                                          & 355k                                                        \\
\rowcolor[HTML]{EFEFEF} 
Sum         &                                                       &                                                         &                                                        &                                                          &                                                       &                                                        &                                                           &                                                       & 151k                                                     & 941k                                                          & 1,784k                                                      \\ \midrule
\multicolumn{12}{l}{{\color[HTML]{9B9B9B} \textit{\textbf{Joint Audio and Speech Datasets}}}}                                                                                                                                                                                                                                                                                                                                                                                                                                                                                                                                                                                   \\ \midrule
AudioSet~\cite{gemmeke2017audio}    & x                                                     & -                                                       & x                                                      & x                                                        & -                                                     & -                                                      & -                                                         & x                                                     & 82k                                                      & -                                                             & 747k                                                        \\
FMA~\cite{defferrard2017fma}         & -                                                     & -                                                       & x                                                      & -                                                        & -                                                     & -                                                      & -                                                         & x                                                     & 93k                                                      & 93k                                                           & 396k                                                        \\
\rowcolor[HTML]{EFEFEF} 
Sum         &                                                       &                                                         &                                                        &                                                          &                                                       &                                                        &                                                           &                                                       & 175k                                                     & 93k                                                           & 1,143k                                                      \\ \midrule
\rowcolor[HTML]{C0C0C0} 
Total       & \multicolumn{8}{c}{\cellcolor[HTML]{C0C0C0}(9,641k Question Answer Pairs)}                                                                                                                                                                                                                                                                                                                                                                                               & 1,089k                                                   & 2,951k                                                        & 6,690k                                                      \\ \bottomrule
\end{tabular}
\end{table*}

\squeezeup\squeezeup
\subsection{Open-Ended AQA Generation}

Generating diverse open-ended AQA pairs at a large scale poses challenges with human-based efforts being impractical, and rule-based methods limiting output diversity. We thus use \emph{Audio Instruction Generation} (AIG) proposed in~\cite{gong2023listen} to generate open-ended AQAs with GPT-3.5-Turbo assistance. Specifically, since GPT does not take audio or speech as input, we input the meta information of the audio (e.g., audio events, speech style, emotion, and spoken text) to the GPT-3.5-Turbo model in the form of pure text as a surrogate, and then use the prompt shown in Table~\ref{tab:sample} to let the GPT model generate AQA pairs. As shown in Table~\ref{tab:sample}, the generated open-ended QA pairs are diverse and of high quality. Notably, 65.1\% of open-ended questions appear only once in the dataset.

Note that AIG is only used for data generation; during model training, only the raw audio and generated QA pairs are input to the \texttt{LTU-AS} model. Thus, the model is forced to learn directly from the raw audio that contains richer and more fine-grained information compared to the extracted meta-information. Similarly, during inference, \texttt{LTU-AS} solely uses raw audio to answer the question.

\textbf{Open-Ended Audio AQA}: We use the approximately 3.7 million AQAs about non-speech audio events from the original OpenAQA dataset~\cite{gong2023listen}.

\textbf{Open-Ended Speech AQA}: We generate open-ended AQAs about speech using the four commonly used datasets IEMOCAP~\cite{busso2008iemocap}, MOSEI~\cite{zadeh2018multimodal}, LibriTTS~\cite{zen2019libritts}, and VoxCeleb2~\cite{chung2018voxceleb2,hechmi2021voxceleb}. We input all speech meta information including the original dataset labels (e.g., speaker emotion, gender, and age), extracted speech style features (e.g., pitch, speed, and volume), and Whisper transcribed spoken text, all in text form, to GPT-3.5-Turbo with the prompt shown in Table~\ref{tab:sample}. For age, pitch, speed, and volume, we also quantize each of them into 5 categories (e.g., very low - very high) to help GPT understand the value. The input meta information to GPT of each dataset is marked as ``x'' in Table~\ref{tab:openasqa}.  Our intent was to input as much information as possible to GPT to generate high-quality AQAs.

\textbf{Open-Ended Joint Audio and Speech AQA} 

We use two datasets containing both speech and non-speech audio to generate joint audio and speech AQAs. The first dataset we use is AudioSet~\cite{gemmeke2017audio}. Although AudioSet-2M has about 1M samples containing speech, and it has already been used in the original OpenAQA dataset, the spoken text was ignored. Specifically, a single label ``speech'' rather than the actual spoken text is input to GPT-3.5-Turbo for OpenAQA generation. In this work, we first sample a 500k subset from AudioSet-2M using the sound class balancing algorithm proposed in~\cite{gong_psla} to guarantee the diversity of non-speech audio events. We then use Whisper to transcribe the 500k AudioSet subset and select samples having \texttt{no\_speech\_prob}$<$0.2 and spoken text length over 5. This heuristic made it quite likely that the spoken text was transcribed correctly and had sufficient length to encompass substantive content. This resulted in 82k samples meeting the requirement.  They were used to generate joint audio and speech AQAs with GPT assistance. As shown in Table~\ref{tab:sample}, GPT can generate AQAs for joint audio and speech understanding, e.g., in the first sample, GPT outputs an answer explaining the speech is addressing the dog by understanding the speech content and the dog sound.

The second dataset we use is the FMA~\cite{defferrard2017fma} music dataset.  We input the list of music genres, title (if provided), and Whisper transcribed lyrics of each music clip to GPT and let it generate AQAs about music understanding with joint lyrics and melody analysis. In total, we generated about 1.1 million open-ended joint audio and speech AQAs.

\begin{table}[h]
\footnotesize
\centering
\squeezeup
\caption{The LTU-AS training curriculum.}
\label{tab:corr}
\setlength\tabcolsep{3pt}
\begin{tabular}{@{}cccccc@{}}
\toprule
\multicolumn{1}{l}{Stage} & Tr. Params          & Tr. Task & Tr. Samples & LR   & \# Epochs \\ \midrule
1                         & Proj.               & Cla.     & 2.1M        & 1e-3 & 2         \\
2                         & Proj. + TLTR + LoRA & Cla.     & 2.1M        & 2e-4 & 2         \\
3                         & Proj. + TLTR + LoRA & All      & 9.6M        & 2e-4 & 1         \\ \bottomrule
\end{tabular}
\squeezeup
\end{table}

\begin{table*}[t]
\footnotesize
\centering
\squeezeup
\caption{Closed-ended task performance. ZS: Zero-shot evaluation; ZS-: The dataset is not used in training, but it is sourced from the same project as part of the training data. * Model does not follow instructions on part of or entire of the dataset. }
\setlength\tabcolsep{5.5pt}
\label{tab:close_res}
\begin{tabular}{@{}lccccccc@{}}
\toprule
                                                      & \begin{tabular}[c]{@{}c@{}}Audio\\ Classif.\end{tabular} & \begin{tabular}[c]{@{}c@{}}Audio\\ Caption\end{tabular}     & \begin{tabular}[c]{@{}c@{}}Speech\\ Recognition\end{tabular} & \begin{tabular}[c]{@{}c@{}}Emotion\\ Recognition\end{tabular} & \begin{tabular}[c]{@{}c@{}}Gender\\ Classif.\end{tabular}      & \begin{tabular}[c]{@{}c@{}}Age\\ Pred.\end{tabular}       & \begin{tabular}[c]{@{}c@{}}Music Genre\\ Classif.\end{tabular} \\ \cmidrule(l){2-8} 
\multirow{-2}{*}{Model}                               & \begin{tabular}[c]{@{}c@{}}ESC-50~\cite{piczak2015esc} \\ (ACC $\uparrow$)\end{tabular}  & \begin{tabular}[c]{@{}c@{}}AudioCaps\\ (SPICE $\uparrow$)\end{tabular} & \begin{tabular}[c]{@{}c@{}}Librispeech~\cite{panayotov2015librispeech}\\ (\texttt{test-clean} WER $\downarrow$)\end{tabular}  & \begin{tabular}[c]{@{}c@{}}IEMOCAP\\ (ACC $\uparrow$)\end{tabular}       & \begin{tabular}[c]{@{}c@{}}Voxceleb2\\ (macro-F1 $\uparrow$)\end{tabular} & \begin{tabular}[c]{@{}c@{}}Voxceleb2\\ (MAE $\downarrow$)\end{tabular} & \begin{tabular}[c]{@{}c@{}}GTZAN~\cite{sturm2013gtzan}\\ (ACC $\uparrow$)\end{tabular}          \\ \midrule
\multicolumn{8}{l}{{\color[HTML]{656565} Best specialized models trained supervisedly on each dataset. Not generalizable to unseen label sets and tasks.}}                                                                                                                                                                                                                                                                                                                                                  \\
{\color[HTML]{656565} Best Supervised \& Specialized} & {\color[HTML]{656565} 97.0~\cite{chen2022hts}}                              & {\color[HTML]{656565} 17.7~\cite{kim2022improving}}                                 & {\color[HTML]{656565} 1.4~\cite{zhang2022bigssl}}                                   & {\color[HTML]{656565} 70.6~\cite{chen2022wavlm}}                                   & {\color[HTML]{656565} 98.3~\cite{hechmi2021voxceleb}}                                    & {\color[HTML]{656565} 9.4~\cite{hechmi2021voxceleb}}                                & {\color[HTML]{656565} 93.9~\cite{liu2021bottom}}                                    \\ \midrule
\multicolumn{8}{l}{{\color[HTML]{656565} CLIP-like audio-text model. Generalizable to unseen labels, but a pre-defined label set is required for inference}}                                                                                                                                                                                                                                                                                                                                                \\
AudioClip~\cite{guzhov2022audioclip}                                             & 69.4                                                     & -                                                           & -                                                            & -                                                             & -                                                              & -                                                         & -                                                              \\
CLAP~\cite{elizalde2023clap}                                                  & 82.6                                                     & -                                                           & -                                                            & -                                                             & -                                                              & -                                                         & 25.2                                                           \\ \midrule
\multicolumn{8}{l}{{\color[HTML]{656565} (Proposed) One single model for all tasks. Directly output label names, no pre-defined label set is needed at inference.}}                                                                                                                                                                                                                                                                                                                                         \\
\rowcolor[HTML]{C0C0C0} 
LTU-AS                                                & 80.8 \textsuperscript{zs-}                                                    & 15.0                                                        & \textbf{4.9}                                                 & 65.2                                                          & 90.8                                                  & \textbf{7.3}                                              & \textbf{50.3}\textsuperscript{zs}                                                  \\ \midrule
\multicolumn{8}{l}{{\color[HTML]{656565} Ablation Study 1 - Train with only speech or audio data}}                                                                                                                                                                                                                                                                                                                                                                                                          \\
LTU (Audio Training Only)~\cite{gong2023listen}                                & \textbf{82.8}                                            & \textbf{17.0}                                               & 104.2                                                        & 38.2                                                          & 77.0                                                           & Fail*                                                      & 29.8                                                           \\
LTU (Speech Training Only)                               & 10.9                                                     & 0.5                                                         & 12.9                                                         & \textbf{69.8}                                                 & 90.1                                                           & 7.9                                                       & 23.5                                                           \\ \midrule
\multicolumn{8}{l}{{\color[HTML]{656565} Ablation Study 2 - Inference with missing modality}}                                                                                                                                                                                                                                                                                                                                                                                                               \\
LTU-AS (Audio Input Only)                             & 81.9                                                     & 14.9                                                        & 97.2                                                         & 58.6                                                          & \textbf{95.6}                                                           & 8.2                                                       & 48.2                                                           \\
LTU-AS (Spoken Text Input Only)                              & 7.7                                                      & 3.5                                                         & 20.0                                                         & 45.4                                                          & 42.0                                                           & 11.9*                                                     & 21.5                                                           \\ \bottomrule
\end{tabular}
\squeezeup\squeezeup
\end{table*}

\squeezeup
\section{Training LTU-AS}
\label{sec:train}
\squeezeup

As for the LTU model~\cite{gong2023listen}, we use a three-stage training curriculum shown in Table~\ref{tab:corr} to train \texttt{LTU-AS}. In the first stage, only the randomly initialized projection layer is trainable. The TLTR and LoRA adapters are unfrozen in the second and third stages to stabilize training. In addition, in the first and second stages, we only train \texttt{LTU-AS} with AQAs of classification tasks where the model gets a high penalty for wrong predictions. The model is thus forced to attend to the audio input rather than using its language ability to hallucinate. 

\section{Experiments}
\label{sec:exp}
\squeezeup

\subsection{Closed-Ended Tasks Evaluation}
\squeezeup
Although the main novelty of \texttt{LTU-AS} is open-ended audio and speech understanding, we first rigorously evaluate its performance on seven standard closed-ended audio and speech tasks because these tasks serve as the foundation for advanced reasoning. Specifically, for each task, we use a fixed prompt (e.g., ``write an audio caption describing the sound.'' for audio classification) and either apply a regular expression to the \texttt{LTU-AS} to get the prediction (for ASR, audio captioning, gender classification, and age prediction), or compute the cosine similarity between the text embedding (gpt-text-embedding-ada) of \texttt{LTU-AS} output and each label, and use label that has the highest similarity score as the prediction (for other classification tasks). 

The results are summarized in Table~\ref{tab:close_res}. First, as a foundational model, \texttt{LTU-AS} performs well on both audio and speech tasks. It works particularly well on tasks requiring both audio and speech understanding, which exactly meets our expectations. E.g., the accuracy of \texttt{LTU-AS} is nearly twice that of CLAP~\cite{elizalde2023clap} on the zero-shot GTZAN music genre classification task; the MAE of speaker age prediction is even lower than the SOTA specialized model that only works for the task. Compared with CLIP-like models~\cite{elizalde2023clap,guzhov2022audioclip}, \texttt{LTU-AS} does not require any pre-defined label set and directly outputs predictions in natural language, which makes it a more practical system for real-world applications.

Second, training with both non-speech audio and speech data is crucial for \texttt{LTU-AS} to become a unified sound perception model. In Ablation Study 1, we compare \texttt{LTU-AS} with LTU models trained with only audio and only speech datasets. Though audio- and speech-specialized LTUs perform slightly better on tasks in their respective training domain, they almost fail on tasks in the domain they are not trained on. 

Third, to take a closer look at how LLaMA attends to continuous audio token input $\{A\}$ and spoken text token input $\{S\}$ on different tasks, we manually remove one input modality for Ablation Study 2. For most tasks, a missing modality leads to a performance drop, indicating that LLaMA takes both $\{A\}$ and $\{S\}$ into its decision-making. Even on audio classification and gender classification tasks where $\{S\}$ is not useful, including $\{S\}$ leads to only a slight performance drop, demonstrating that \texttt{LTU-AS} can correctly attend to $\{A\}$ and $\{S\}$ based on the input audio and question. Finally, we observe the ASR performance of \texttt{LTU-AS} (4.9\% WER) is worse than its internal Whisper model (3.5\% WER) due to occasionally not following instructions and changing spelling.

\begin{table}[h]
\footnotesize
\centering
\squeezeup\squeezeup
\caption{LTU instruction following rate rated by GPT-4.}
\label{tab:ins}
\begin{tabular}{@{}lcc@{}}
\toprule
Model  & Audio Question & Speech Question \\ \midrule
LTU-Audio Training Only  & 96\%           & 69\%            \\
LTU-Speech Training Only  & 65\%           & 93\%            \\
LTU-AS & \textbf{96\%}           & \textbf{94\%}            \\ \bottomrule
\end{tabular}
\squeezeup\squeezeup\squeezeup\squeezeup
\end{table}

\begin{table}
\footnotesize
\squeezeup\squeezeup
\caption{Real samples of \texttt{LTU-AS} open-ended question answering on evaluation samples. \gt{Acoustic} and \rt{speech} information in the answers are marked in \gt{green} and \rt{red}, respectively, to highlight how \texttt{LTU-AS} understand them jointly.}
\label{tab:open_res}
\begin{tabular}{p{8.0cm}}
\hline
Sample 1 Basketball Coaching (AS-Eval)                                                            \\ 
\grt{Transcribed speech: To the left, the second time through they'll do the same move to the right. We'll do a series of about five moves. Here we've got a 6'9 kid, I expect him to make the same move.}                                                                                       \\ \hline
\bt{Q: What can be inferred from the audio events and spoken text?}                               \\
A: \gt{The audio event is a basketball bounce} while \rt{the spoken text suggests that someone is giving instructions}.                                                                             \\
\bt{Q: what is the job of the speaker?}                                                           \\
A: The speaker is \rt{coaching or instructing someone on how to play basketball based on their speech content} and \gt{the background sounds of basketball bounce heard}.                                     \\
\bt{Q: Where could this audio be recorded?}                                                       \\
A: This audio clip could have been recorded in a basketball court or gymnasium where \gt{people are playing basketball} and \rt{speaking}.                                                                  \\ \hline
Sample 2 Interview (VoxCeleb2-Eval)                                                               \\ 
\grt{Transcribed speech: ... I was so anxious about getting up and talking in front of people that I thought I still sort of lose...}                                                           \\ \hline
\bt{Q: What is the emotion of the speaker? Why?}                                                  \\
A: \rt{Based on the speech content, the speaker seems to express anxiety or nervousness about public speaking}. \gt{The high pitch and fast speed suggest excitement or tension as well.} Therefore, the emotion of the speaker can be inferred as nervousness or anxiety.                                 \\
\bt{Q: Is the speaker a good public speaker?}                                                     \\
A: \rt{No, the speaker feels anxious about speaking in front of people and thinks they will still feel lose-lose about it}.                                                                          \\
\bt{Q: Provide the speaker 5 suggestions.}                                                        \\
A: \rt{1. Take a deep breath before speaking. 2. Practice public speaking regularly. 3. Use humor or lightheartedness to ease nerves. 4. Focus on the audience rather than yourself during speeches. 5. Seek feedback from others to improve your delivery.}                                           \\ \hline
Sample 3 Rock Music (GTZAN)                                                                       \\ 
\grt{Transcribed Lyrics: Gotta be tonight, my days are blue, babe Mama says you don't, daddy says you won't. And I'm falling off her side No way I'm gonna lose out this time. Tie your mother down...} \\ \hline
\bt{Q: What did you hear?} \gt{A: Music, punk rock.}                                              \\
\bt{Q: Why the music is rock? Consider both lyrics and sounds.}                                   \\
A: The music has \gt{a strong beat, distorted guitar riffs, and aggressive vocals that are typical characteristics of rock music}. \rt{The lyrics also contain themes of rebellion and defiance which align with the genre's cultural significance.}                                                     \\
\bt{Q: Can you suggest a title for this music?}                                                   \\
A: \rt{Tie Your Mother Down} - \gt{A Punk Rock Anthem.}                                                     \\
\bt{Q: Is this music appropriate to play in a primary school?}                                    \\
A: No, \rt{the lyrics} and \gt{overall tone} are not suitable for children.                                 \\
\bt{Q: What mood does this music convey?}                                                         \\
A: The music conveys a rebellious and energetic mood with \gt{its fast-paced tempo and aggressive guitar riffs}. \rt{The lyrics also add to the sense of rebellion.}                                \\ \hline
Sample 4 Subway Station (FSD50K)                                                                  \\ 
\grt{Transcribed speech: This is the transfer point for the Metro Purple Line to Wilshire Western. Boarding platform is located downstairs...}                                                \\ \hline
\bt{Q: What did you hear?} A: \gt{Subway, metro, underground.}                                    \\
\bt{Q: What spoken text is heard?} \\ A: The spoken text is \rt{``This is the transfer point for the Metro Purple Line to Wilshire Western. Boarding platform is located downstairs...''}                                    \\
\bt{Q: What can be inferred from the sound and announcement?}                                     \\
A: The \gt{subway sound suggests that it is a busy station with many people coming and going}, while \rt{the announcement provides information about the train schedule.}                        \\
\bt{Q: Where is the boarding platform located?}                                                   \\
A: \rt{The boarding platform is located downstairs.}                                              \\
\bt{Q: Which line does this station transfer to?}                                                 \\
A: \rt{The transfer point is to the Metro Purple Line.}                                           \\
\bt{Q: What should I do if I hear this?}                                                          \\
A: \rt{You should board the next train and proceed to your destination.}                          \\ \hline
\end{tabular}
\end{table}

\subsection{Open-Ended Audio Question Answering Evaluation}

In addition to the good performance on closed-ended tasks, \texttt{LTU-AS} also exhibits superior performance when it comes to answering open-ended questions. We quantitatively measure the instruction following rate of \texttt{LTU-AS} on audio and speech questions and compare it with LTU models trained with only audio data and speech data. Specifically, we use GPT-4 to generate 100 audio and speech questions based on AudioSet and VoxCeleb evaluation sets, respectively, and input the questions and corresponding audios to LTU models and collect its answers. Finally, we use GPT-4 to evaluate if the LTU model output answers the given questions by the prompt ``Below is a pair of question and response. Identify if the response directly answers the question and give a clear answer.'' As shown in Table~\ref{tab:ins}, \texttt{LTU-AS} has an instruction following rate over 94\% for both audio and speech questions, while LTU trained with only audio/speech dataset does not follow instruction well on questions out of its training domain. Please note that this automatic evaluation protocol may overestimate the actual instruction following rate because 1) the evaluation questions are generated by GPT-4, not humans; and 2) GPT-4 may overrate \texttt{LTU-AS} answers as \texttt{LTU-AS} is trained to generate answers similar to GPT outputs.

As shown in Table~\ref{tab:open_res}, \textbf{LTU-AS understands the world by combing audio and speech information}: In example 1, \texttt{LTU-AS} correctly identifies the job of the speaker as a basketball coach because the spoken text is about instructing while bouncing basketballs are heard in the background. Without understanding the spoken text, the speaker could be a basketball player, while without understanding the audio, the speaker could be a football coach. Similarly, in example 2, \texttt{LTU-AS} knows the speaker is anxious because of spoken content and expresses concern about public speaking while the speaker speaks fast with a high pitch. \textbf{LTU-AS exhibits emerging reasoning ability and connects sounds to actions}: In Sample 2, \texttt{LTU-AS} can provide suggestions to the speaker based on his situation; in Sample 3, \texttt{LTU-AS} can suggest a title for the music, and does not recommend to play it in a primary school because the lyrics and music tone are not suitable for children; in Sample 4, \texttt{LTU-AS} not only correctly extracts the information about the boarding platform and transfer line, but also suggests boarding the next train when we hear the announcement. All these demonstrate \texttt{LTU-AS} can comprehend the input audio and speech.

\squeezeup\squeezeup
\section{Conclusions}
\squeezeup\squeezeup

In this paper, we present \texttt{LTU-AS}, a novel joint audio and speech understanding model that can simultaneously recognize and jointly understand spoken text, speech paralinguistics, and non-speech audio events. We identify three key components in successfully building \texttt{LTU-AS}. First, \texttt{LTU-AS} uses a strong audio encoder (Whisper) and a strong reasoning model (LLaMA). The former provides precise perception ability while the latter provides advanced reasoning ability. Second, \texttt{LTU-AS} is trained with the new large-scale dataset \texttt{Open-ASQA} with a wide range of diverse audio and speech tasks. In particular, the open-ended questions generated with GPT assistance are crucial to empower \texttt{LTU-AS} to answer free-form questions. Third, \texttt{LTU-AS} is trained with a multi-stage training curriculum to militate hallucination. \texttt{LTU-AS} achieves good performance on all tested closed-ended audio and speech benchmarks, particularly on tasks requiring joint audio and speech understanding. More importantly, when answering free-form, open-ended questions, \texttt{LTU-AS} effectively combines information from audio and speech, and exhibits emerging reasoning abilities.

\vspace{0mm}
\noindent{\textbf{Ethics Statement}}: To prevent the potential misuse of the proposed audio LLM, we intentionally do not include speaker identification in our task list. The music used in model training is Creative Commons-licensed~\cite{defferrard2017fma}.

\bibliographystyle{IEEEbib}
\bibliography{refs}

\begin{thebibliography}{10}

\bibitem{radford2022robust}
Alec Radford, Jong~Wook Kim, Tao Xu, Greg Brockman, Christine McLeavey, and Ilya Sutskever,
\newblock ``{Robust Speech Recognition via Large-scale Weak Supervision},''
\newblock {\em arXiv preprint arXiv:2212.04356}, 2022.

\bibitem{touvron2023llama}
Hugo Touvron, Thibaut Lavril, Gautier Izacard, Xavier Martinet, Marie-Anne Lachaux, Timoth{\'e}e Lacroix, Baptiste Rozi{\`e}re, Naman Goyal, Eric Hambro, Faisal Azhar, et~al.,
\newblock ``{LLaMA: Open and Efficient Foundation Language Models},''
\newblock {\em arXiv preprint arXiv:2302.13971}, 2023.

\bibitem{gong2023whisper}
Yuan Gong, Sameer Khurana, Leonid Karlinsky, and James Glass,
\newblock ``{Whisper-AT: Noise-Robust Automatic Speech Recognizers are Also Strong General Audio Event Taggers},''
\newblock in {\em Interspeech}, 2023.

\bibitem{brown2020language}
Tom Brown, Benjamin Mann, Nick Ryder, Melanie Subbiah, Jared~D Kaplan, Prafulla Dhariwal, Arvind Neelakantan, Pranav Shyam, Girish Sastry, Amanda Askell, et~al.,
\newblock ``{Language Models are Few-shot Learners},''
\newblock {\em NeurIPS}, 2020.

\bibitem{gong2023listen}
Yuan Gong, Hongyin Luo, Alexander~H Liu, Leonid Karlinsky, and James Glass,
\newblock ``{Listen, Think, and Understand},''
\newblock {\em arXiv preprint arXiv:2305.10790}, 2023.

\bibitem{gong21b_interspeech}
Yuan Gong, Yu-An Chung, and James Glass,
\newblock ``{AST: Audio Spectrogram Transformer},''
\newblock in {\em Interspeech}, 2021.

\bibitem{dinkel2022unikw}
Heinrich Dinkel, Yongqing Wang, Zhiyong Yan, Junbo Zhang, and Yujun Wang,
\newblock ``{Unified Keyword Spotting and Audio Tagging on Mobile Devices with Transformers},''
\newblock in {\em ICASSP}, 2023.

\bibitem{moritz2020all}
Niko Moritz, Gordon Wichern, Takaaki Hori, and Jonathan Le~Roux,
\newblock ``{All-in-One Transformer: Unifying Speech Recognition, Audio Tagging, and Event Detection},''
\newblock in {\em Interspeech}, 2020.

\bibitem{narisetty2022joint}
Chaitanya Narisetty, Emiru Tsunoo, Xuankai Chang, Yosuke Kashiwagi, Michael Hentschel, and Shinji Watanabe,
\newblock ``{Joint Speech Recognition and Audio Captioning},''
\newblock in {\em ICASSP}, 2022.

\bibitem{turian2022hear}
Joseph Turian, Jordie Shier, Humair~Raj Khan, et~al.,
\newblock ``{HEAR: Holistic Evaluation of Audio Representations},''
\newblock in {\em NeurIPS Competitions and Demonstrations Track}, 2021.

\bibitem{gao2022wavprompt}
Heting Gao, Junrui Ni, Kaizhi Qian, Yang Zhang, Shiyu Chang, and Mark Hasegawa-Johnson,
\newblock ``{WavPrompt: Towards Few-Shot Spoken Language Understanding with Frozen Language Models},''
\newblock in {\em Interspeech}, 2022.

\bibitem{huang2023audiogpt}
Rongjie Huang, Mingze Li, Dongchao Yang, Jiatong Shi, Xuankai Chang, Zhenhui Ye, Yuning Wu, Zhiqing Hong, Jiawei Huang, Jinglin Liu, et~al.,
\newblock ``{AudioGPT: Understanding and Generating Speech, Music, Sound, and Talking Head},''
\newblock {\em arXiv preprint arXiv:2304.12995}, 2023.

\bibitem{chang2023speechprompt}
Kai-Wei Chang, Yu-Kai Wang, Hua Shen, et~al.,
\newblock ``{SpeechPrompt v2: Prompt Tuning for Speech Classification Tasks},''
\newblock {\em arXiv preprint arXiv:2303.00733}, 2023.

\bibitem{zhang2023speechgpt}
Dong Zhang, Shimin Li, Xin Zhang, Jun Zhan, Pengyu Wang, Yaqian Zhou, and Xipeng Qiu,
\newblock ``{SpeechGPT: Empowering Large Language Models with Intrinsic Cross-modal Conversational Abilities},''
\newblock {\em arXiv preprint arXiv:2305.11000}, 2023.

\bibitem{feng2023peft}
Tiantian Feng and Shrikanth Narayanan,
\newblock ``{PEFT-SER: On the Use of Parameter Efficient Transfer Learning Approaches For Speech Emotion Recognition Using Pre-trained Speech Models},''
\newblock {\em arXiv preprint arXiv:2306.05350}, 2023.

\bibitem{xu2023understanding}
Anfeng Xu, Rajat Hebbar, Rimita Lahiri, Tiantian Feng, Lindsay Butler, Lue Shen, Helen Tager-Flusberg, and Shrikanth Narayanan,
\newblock ``{Understanding Spoken Language Development of Children with ASD Using Pre-trained Speech Embeddings},''
\newblock in {\em Interspeech}, 2023.

\bibitem{vaswani2017attention}
Ashish Vaswani, Noam Shazeer, Niki Parmar, Jakob Uszkoreit, Llion Jones, Aidan~N. Gomez, Lukasz Kaiser, and Illia Polosukhin,
\newblock ``{Attention is All You Need},''
\newblock in {\em NeurIPS}, 2017.

\bibitem{vicuna2023}
Wei-Lin Chiang, Zhuohan Li, Zi~Lin, et~al.,
\newblock ``{Vicuna: An Open-Source Chatbot Impressing GPT-4 with 90\%* ChatGPT Quality},'' March 2023.

\bibitem{goodfellow2013empirical}
Ian~J Goodfellow, Mehdi Mirza, Da~Xiao, Aaron Courville, and Yoshua Bengio,
\newblock ``{An Empirical Investigation of Catastrophic Forgetting in Gradient-based Neural Networks},''
\newblock {\em arXiv preprint arXiv:1312.6211}, 2013.

\bibitem{hu2021lora}
Edward~J Hu, Yelong Shen, Phillip Wallis, Zeyuan Allen-Zhu, Yuanzhi Li, Shean Wang, Lu~Wang, and Weizhu Chen,
\newblock ``{LoRA: Low-rank Adaptation of Large Language Models},''
\newblock {\em arXiv preprint arXiv:2106.09685}, 2021.

\bibitem{fan2018hierarchical}
Angela Fan, Mike Lewis, and Yann Dauphin,
\newblock ``{Hierarchical Neural Story Generation},''
\newblock in {\em ACL}, 2018.

\bibitem{keskar2019ctrl}
Nitish~Shirish Keskar, Bryan McCann, Lav~R Varshney, Caiming Xiong, and Richard Socher,
\newblock ``{Ctrl: A Conditional Transformer Language Model for Controllable Generation},''
\newblock {\em arXiv preprint arXiv:1909.05858}, 2019.

\bibitem{busso2008iemocap}
Carlos Busso, Murtaza Bulut, Chi-Chun Lee, et~al.,
\newblock ``{IEMOCAP: Interactive Emotional Dyadic Motion Capture Database},''
\newblock {\em Language resources and evaluation}, 2008.

\bibitem{zadeh2018multimodal}
AmirAli~Bagher Zadeh, Paul~Pu Liang, Soujanya Poria, Erik Cambria, and Louis-Philippe Morency,
\newblock ``{Multimodal Language Analysis in the Wild: CMU-MOSEI Dataset and Interpretable Dynamic Fusion Graph},''
\newblock in {\em ACL}, 2018.

\bibitem{zen2019libritts}
Heiga Zen, Viet Dang, Rob Clark, Yu~Zhang, Ron~J Weiss, Ye~Jia, Zhifeng Chen, and Yonghui Wu,
\newblock ``{LibriTTS: A Corpus Derived from LibriSpeech for Text-to-Speech},''
\newblock {\em arXiv preprint arXiv:1904.02882}, 2019.

\bibitem{chung2018voxceleb2}
Joon~Son Chung, Arsha Nagrani, and Andrew Zisserman,
\newblock ``{VoxCeleb2: Deep Speaker Recognition},''
\newblock {\em arXiv preprint arXiv:1806.05622}, 2018.

\bibitem{hechmi2021voxceleb}
Khaled Hechmi, Trung~Ngo Trong, Ville Hautam{\"a}ki, and Tomi Kinnunen,
\newblock ``{VoxCeleb Enrichment for Age and Gender Recognition},''
\newblock in {\em ASRU}, 2021.

\bibitem{mcfee2015librosa}
Brian McFee, Colin Raffel, Dawen Liang, Daniel~P Ellis, Matt McVicar, Eric Battenberg, and Oriol Nieto,
\newblock ``{librosa: Audio and Music Signal Analysis in Python},''
\newblock in {\em The 14th Python in Science Conference}, 2015.

\bibitem{defferrard2017fma}
Micha{\"e}l Defferrard, Kirell Benzi, Pierre Vandergheynst, and Xavier Bresson,
\newblock ``{FMA: A Dataset For Music Analysis},''
\newblock in {\em 18th International Society for Music Information Retrieval Conference}, 2017.

\bibitem{hershey2021benefit}
Shawn Hershey, Daniel~PW Ellis, Eduardo Fonseca, Aren Jansen, Caroline Liu, R~Channing Moore, and Manoj Plakal,
\newblock ``{The Benefit Of Temporally-Strong Labels In Audio Event Classification},''
\newblock in {\em ICASSP}, 2021.

\bibitem{gemmeke2017audio}
Jort~F Gemmeke, Daniel~PW Ellis, Dylan Freedman, Aren Jansen, Wade Lawrence, R~Channing Moore, Manoj Plakal, and Marvin Ritter,
\newblock ``{Audio Set: An Ontology and Human-labeled Dataset for Audio Events},''
\newblock in {\em ICASSP}, 2017.

\bibitem{chen2020vggsound}
Honglie Chen, Weidi Xie, Andrea Vedaldi, and Andrew Zisserman,
\newblock ``{VGGSound: A Large-scale Audio-Visual Dataset},''
\newblock in {\em ICASSP}, 2020.

\bibitem{fonseca2021fsd50k}
Eduardo Fonseca, Xavier Favory, Jordi Pons, Frederic Font, and Xavier Serra,
\newblock ``{FSD50K: An Open Dataset of Human-Labeled Sound Events},''
\newblock {\em IEEE Transactions on Audio, Speech, and Language Processing}, 2021.

\bibitem{kim2019audiocaps}
Chris~Dongjoo Kim, Byeongchang Kim, Hyunmin Lee, and Gunhee Kim,
\newblock ``{AudioCaps: Generating Captions for Audios in The Wild},''
\newblock in {\em NAACL}, 2019.

\bibitem{freesound}
``{FreeSound Project},'' \url{https://freesound.org/}.

\bibitem{Drossos_2019_dcase}
Samuel Lipping, Konstantinos Drossos, and Tuoams Virtanen,
\newblock ``{Crowdsourcing a Dataset of Audio Captions},''
\newblock in {\em DCASE}, 2019.

\bibitem{soundbible}
``{Sound Bible},'' \url{https://soundbible.com/}.

\bibitem{gong_psla}
Yuan Gong, Yu-An Chung, and James Glass,
\newblock ``{PSLA: Improving Audio Tagging with Pretraining, Sampling, Labeling, and Aggregation},''
\newblock {\em IEEE Transactions on Audio, Speech, and Language Processing}, 2021.

\bibitem{piczak2015esc}
Karol~J Piczak,
\newblock ``{ESC: Dataset for Environmental Sound Classification},''
\newblock in {\em ACM-MM}, 2015.

\bibitem{panayotov2015librispeech}
Vassil Panayotov, Guoguo Chen, Daniel Povey, and Sanjeev Khudanpur,
\newblock ``{Librispeech: An ASR Corpus Based on Public Domain Audio Books},''
\newblock in {\em ICASSP}, 2015.

\bibitem{sturm2013gtzan}
Bob~L Sturm,
\newblock ``{The GTZAN Dataset: Its Contents, Its Faults, Their Effects on Evaluation, and Its Future Use},''
\newblock {\em arXiv preprint arXiv:1306.1461}, 2013.

\bibitem{chen2022hts}
Ke~Chen, Xingjian Du, Bilei Zhu, Zejun Ma, Taylor Berg-Kirkpatrick, and Shlomo Dubnov,
\newblock ``{HTS-AT: A Hierarchical Token-Semantic Audio Transformer for Sound Classification and Detection},''
\newblock in {\em ICASSP}, 2022.

\bibitem{kim2022improving}
Eungbeom Kim, Jinhee Kim, Yoori Oh, Kyungsu Kim, Minju Park, Jaeheon Sim, Jinwoo Lee, and Kyogu Lee,
\newblock ``{Improving Audio-Language Learning with MixGen and Multi-Level Test-Time Augmentation},''
\newblock {\em arXiv preprint arXiv:2210.17143}, 2022.

\bibitem{zhang2022bigssl}
Yu~Zhang, Daniel~S Park, Wei Han, James Qin, Anmol Gulati, Joel Shor, Aren Jansen, Yuanzhong Xu, Yanping Huang, Shibo Wang, et~al.,
\newblock ``{BigSSL: Exploring the Frontier of Large-scale Semi-supervised Learning for Automatic Speech Recognition},''
\newblock {\em IEEE Journal of Selected Topics in Signal Processing}, 2022.

\bibitem{chen2022wavlm}
Sanyuan Chen, Chengyi Wang, Zhengyang Chen, Yu~Wu, Shujie Liu, Zhuo Chen, Jinyu Li, Naoyuki Kanda, Takuya Yoshioka, Xiong Xiao, et~al.,
\newblock ``{WavLM: Large-scale Self-supervised Pre-training for Full Stack Speech Processing},''
\newblock {\em IEEE Journal of Selected Topics in Signal Processing}, 2022.

\bibitem{liu2021bottom}
Caifeng Liu, Lin Feng, Guochao Liu, Huibing Wang, and Shenglan Liu,
\newblock ``{Bottom-up Broadcast Neural Network for Music Genre Classification},''
\newblock {\em Multimedia Tools and Applications}, 2021.

\bibitem{guzhov2022audioclip}
Andrey Guzhov, Federico Raue, J{\"o}rn Hees, and Andreas Dengel,
\newblock ``{AudioCLIP: Extending CLIP to Image, Text and Audio},''
\newblock in {\em ICASSP}, 2022.

\bibitem{elizalde2023clap}
Benjamin Elizalde, Soham Deshmukh, Mahmoud Al~Ismail, and Huaming Wang,
\newblock ``{CLAP: Learning Audio Concepts from Natural Language Supervision},''
\newblock in {\em ICASSP}, 2023.

\end{thebibliography}

\end{document}